# Photometric Identification of Objects from Galaxy Evolution Explorer Survey and Sloan Digital Sky Survey

*K. Preethi[1], S. B. Gudennavar[1], S. G. Bubbly[1], Jayant Murthy[2] and Noah Brosch[3]*
*[1] Department of Physics, Christ University, Bangalore-560 029, Karnataka, India*
*[2] Indian Institute of Astrophysics, II Block, Koramangala, Bangalore-560 034, Karnataka, India*
*[3] The Wise Observatory and the School of Physics and Astronomy, the Raymond and Beverly Sackler Faculty of Exact Sciences, Tel Aviv University, Tel Aviv 69978, Israel*

**ABSTRACT**

We have used *GALEX* and SDSS observations to extract 7 band photometric magnitudes for over 80,000 objects in the vicinity of the North Galactic Pole. Although these had been identified as stars by the SDSS pipeline, we found through fitting with model SEDs that most were, in fact, of extragalactic origin. Only about 9% of these objects turned out to be main sequence stars and about 11% were white dwarfs and red giants collectively, while galaxies and quasars contributed to the remaining 80% of the data. We have classified these objects into different spectral types (for the stars) and into different galactic types (for the galaxies). As part of our fitting procedure, we derive the distance and extinction to each object and the photometric redshift towards galaxies and quasars. This method easily allows for the addition of any number of observations to cover a more diverse range of wavelengths, as well as the addition of any number of model templates. The primary objective of this work is to eventually derive a 3-dimensional extinction map of the Milky Way galaxy.

## 1 INTRODUCTION

Photometric surveys have been an important tool in the study of the structure and evolutionary history of our Galaxy (Bahcall 1986; Majewski 1993). One of the largest surveys to date has been the Sloan Digital Sky Survey (SDSS: York et al. 2000) which obtained 5 band photometry from the blue to the near-IR. SDSS has observed over 35% of the sky and has yielded ground-breaking insights in every field of astronomy (Eisenstein et al. 2011).

The Galaxy Evolution Explorer (*GALEX*: Martin et al. 2005) has observed close to 80% of the sky in two ultraviolet (UV) bands and there has since been an explosion of work focusing on the statistical properties of galaxies with both *GALEX* and SDSS observations. For example, Hutchings & Bianchi (2010) have found that quasars in the redshift range 0.5 – 1.5 could be easily separated from stars and galaxies through their Ly α emission, which manifests itself in the *GALEX* UV bands; that is, a star-free set of quasars could be defined. However, the reverse is not true in that a galaxy-free set of stars cannot be found through standard colour plots.

We have begun a systematic survey of all sources detected by both *GALEX* and SDSS around the North Galactic Pole (NGP). Rather than a colour based approach, which cannot effectively distinguish between stars and galaxies, we have adopted a model-based approach where we fit spectral energy distributions to the observed *GALEX* and SDSS photometric bands. This has allowed us to separate and classify stars and galaxies and to find the extinction in the line of sight. We plan to extend this approach to the entire data set to derive a Galactic extinction map; to probe the stellar distribution; and to identify and classify point sources in the survey.

---

[1] Effective wavelengths from Morrissey et al. (2007) for *GALEX* bands and Fukugita et al. (1996) for SDSS bands.

| Band | Limiting magnitude | Effective Wavelength (Å)[1] |
|---|---|---|
| FUV (AIS) | 19.9 | 1538.6 |
| NUV (AIS) | 20.8 | 2315.7 |
| FUV (MIS) | 22.6 | 1538.6 |
| NUV (MIS) | 22.7 | 2315.7 |
| FUV (DIS) | 24.8 | 1538.6 |
| NUV (DIS) | 24.4 | 2315.7 |
| *u* | 22.0 | 3551 |
| *g* | 22.2 | 4686 |
| *r* | 22.2 | 6165 |
| *i* | 21.3 | 7481 |
| *z* | 20.5 | 8931 |

Table 1. Sensitivity of *GALEX* and SDSS bands. AIS, MIS and DIS are, respectively, the All-sky Imaging Survey, the Medium Imaging Survey and the Deep Imaging Survey.

## 2 DATA

The Galaxy Evolution Explorer (*GALEX*) was launched on April 28, 2003 to observe nearby galaxies but has since observed close to 80% of the sky. The mission was described by Martin et al. (2005) and the data products by Morrissey et al. (2007). An image of the sky (FOV: 1.2°) was produced in each of the two bands (FUV: 1528 Å; NUV: 2271 Å) with a spatial resolution of 5″ to 8″. Most observations were from the All-Sky Imaging Survey (AIS) with integration times of about 100 s but there were a significant number with exposure times of several thousand seconds to as much as 30,000 s (Morrissey et al. 2007). The mission's last complete data release was the General Release 6 (GR6) in 2010 with a final data release expected in the future comprised of all observations before the final termination of the *GALEX* project. The FUV camera was intermittently off due to power problems and was shut down permanently in May 2009; thus many observations only include NUV data. We have used the merged point source catalogue of each observation in this work, where the point sources were extracted using Source Extractor (Bertin & Arnouts 1996) with limiting magnitudes and effective wavelengths in Table 1.

The Sloan Digital Sky Survey (SDSS) uses a 2.5 m wide-angle optical telescope at the Apache Point Observatory for multi-filter imaging and spectroscopic redshift surveys (York et al. 2000). There are 5 photometric bands with effective wavelengths and limiting magnitudes listed in Table 1. We have used the data from Data Release 8 (DR8) covering all imaging data taken up to January 2010, comprising over 14,000 square degrees of sky (Aihara et al. 2011) with 469,053,874 sources extracted by an automated pipeline as described by Lupton et al. (2002).

The combination of the SDSS (DR8) and *GALEX* (GR6) data provide photometric magnitudes for objects over approximately one third of the sky in 7 bands from the FUV (1528 Å) through the NIR (8931 Å). The SDSS group at the Johns Hopkins University have cross-matched sources from the two catalogues using techniques described by Budavari et al. (2009) and the merged catalogue is available on the CasJobs[2] server. The source matching was based entirely on spatial coincidence within the 4" limit of the *GALEX* positional accuracy and thus there were sometimes multiple identifications for a single source. We have used a set of custom SQL and IDL queries on the merged SDSS-*GALEX* catalogue to extract broad-band photometry for all objects with photometric observations by both *GALEX* and SDSS. We have rejected all sources with either duplicate *GALEX* IDs or duplicate SDSS IDs to ensure that the broad-band photometry was not contaminated with false matches. Most of the remaining matches were identified as galaxies by the SDSS pipeline (Lupton et al. 2002) with a much smaller number classified as stars (type = 6 in the catalogue), primarily on the basis of their point spread function (PSF). We have restricted ourselves to those objects identified as point sources with galactic latitudes greater than 75° (NGP) in this work, a total of 84,649 objects.

The distribution of these objects is plotted in a polar plot in Fig. 1 and reflects the *GALEX* observation strategy. Most of the observations were part of the All-sky Imaging Survey (AIS) with typical observation times of about 100 seconds and limiting AB magnitudes of 20.5. A few observations were much deeper (exposure time > 1500 s) with higher limiting magnitudes and these are reflected in Fig. 1 as dark patches with a higher source density. Thus the distribution of sources in Fig. 1 is a consequence of the observation strategy rather than the actual stellar distribution.

---

[2] http://casjobs.SDSS.org/CasJobs/

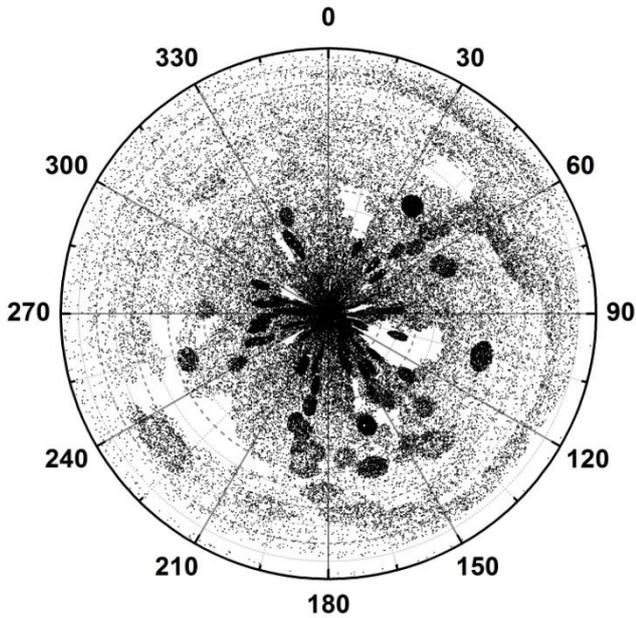

Figure 1. Polar distribution of stars.

## 3 ANALYSIS

We have obtained 7 band photometry for over 80,000 objects around the North Galactic Pole. We have created a grid of model SEDs for stars of different spectral types which were applied to the data reddened by interstellar dust. The stellar models were taken from Castelli & Kurucz (2004) who have published a set of models for different surface temperatures, surface gravities and metallicities. We have restricted ourselves to main sequence stars for which we have used the model parameters listed in the table taken from the documentation of the Castelli & Kurucz models[3]. The gravity (log g) of these models range from 4.0 to 5.0, which correspond to the typical values observed for main sequence stars (Angelov 1996), with a metallicity of log Z = 0.0 and temperatures from 3,500 K to 50,000 K.

The extinction curve was taken from Weingartner & Draine (2001) who have published a set of cross-sections for different reddening curves. We have used their "Milky Way" dust model which is representative of extinction within our Galaxy. Finally, the stellar spectra were convolved with the extinction curve and with filter response curves for *GALEX* (Morrissey et al. 2005) and SDSS (Gunn et al. 1998) to predict band-averaged fluxes for different spectral types and reddening. We used a least-squares procedure to fit the model, with the spectral type of the star, *E(B-V)* and the distance of the star as free parameters, to the data. Individual error bars for each data point were calculated by the SDSS and *GALEX* extraction software and are typically about 0.1 mag for the *GALEX* bands (Morrissey et al. 2007) and about 0.005 mag

---

[3] http://www.stsci.edu/hst/observatory/cdbs/castelli_kurucz_atlas.html

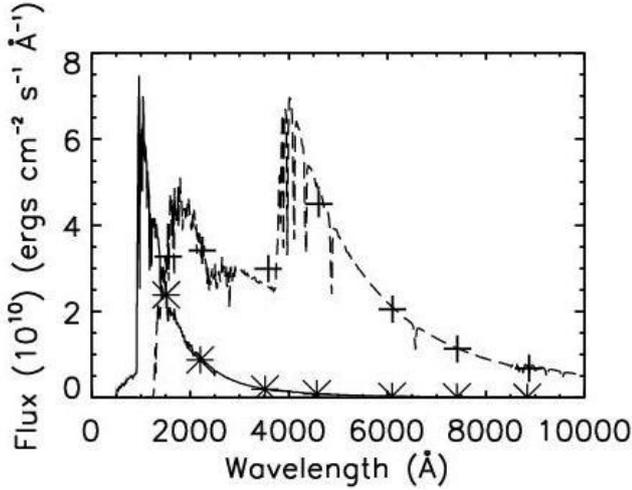

Figure 2. Stellar spectrum of an unreddened A0V star of temperature 9,500 K (dashed line, flux multiplied by a factor of 500) and an unreddened B0V star of temperature 30,000 K (solid line). Cross and asterisk correspond to the effective flux in the 2 *GALEX* and 5 SDSS bands, for the A0V and B0V star respectively.

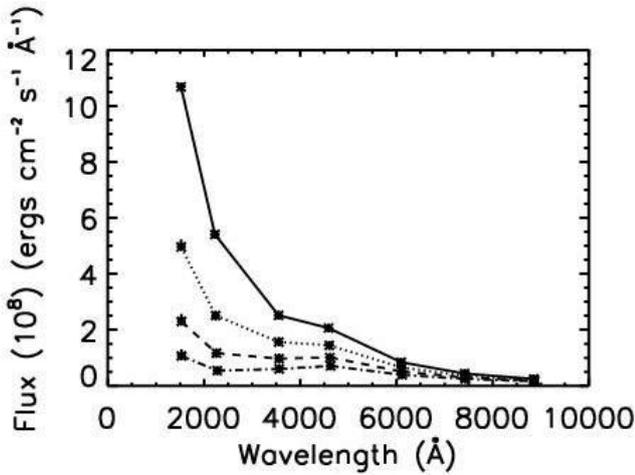

Figure 3. The effective flux for a B0I star through different amounts of reddening, i.e. E(B-V) (solid line); E(B-V) = 0.1 (dotted line); E(B-V) = 0.2 (dashed line) and E(B-V) = 0.3 (dash-dot line).

for each of the SDSS bands.

We have plotted the spectrum and the band averaged fluxes for two stars of different spectral types in Fig. 2 and for a star of a single spectral type with different amounts of reddening in Fig. 3. Using Monte

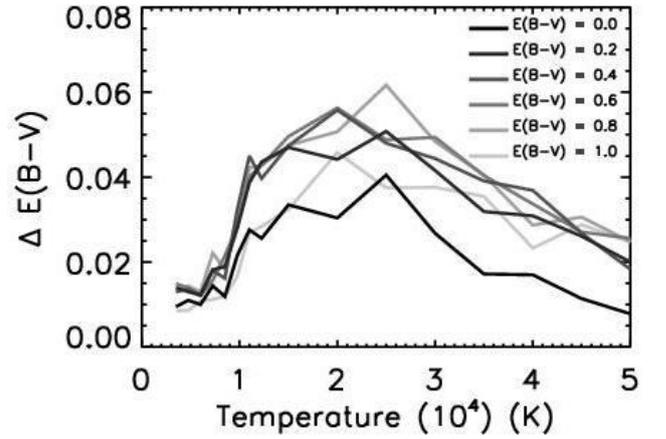

Figure 4. Standard deviations in the *E(B-V)* obtained from Monte Carlo simulations.

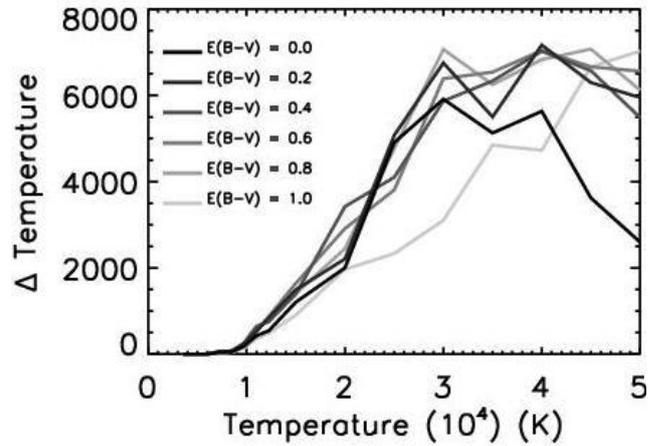

Figure 5. Standard deviations in the temperature obtained from Monte Carlo simulations.

Carlo simulations (see Ravichandran et al. 2013 for a full description) for various combinations of temperature and *E(B-V)* with typical errors in each band, we found that the uncertainty in the derived E(B-V) ranged from 0.02 magnitudes in the case of A type stars to an upper limit of 0.06 magnitudes for B type stars (Fig. 4) while the respective uncertainty in temperature ranged from about 500 K to 7,000 K (Fig. 5). Various such runs, with changes in band errors showed the spectral type

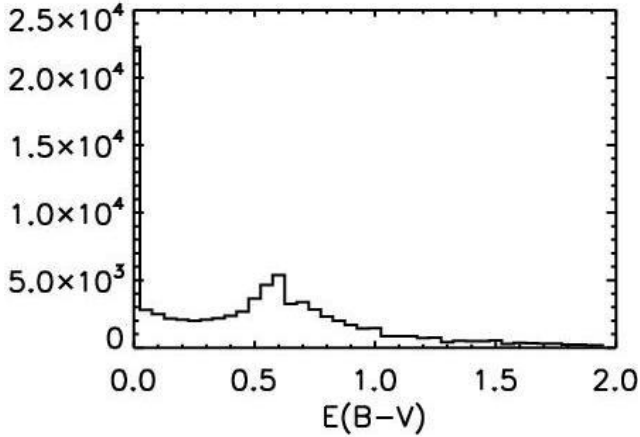

Figure 6. Histogram of best-fitting parameter of extinction.

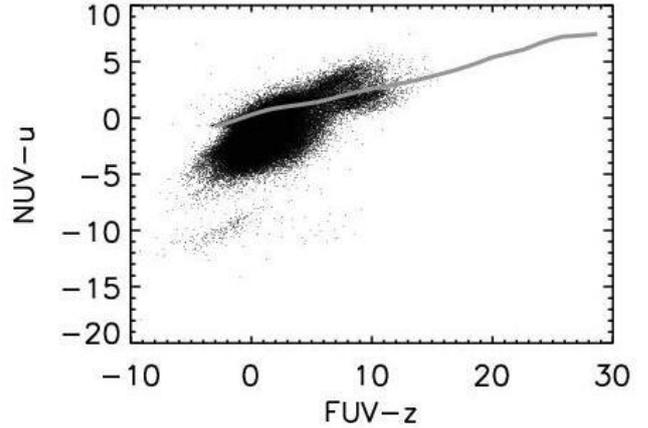

Figure 7. Colour-colour plot of NUV−u vs FUV−z. Objects that don't fall in the same colour-colour space as the Castelli & Kurucz stellar models (unreddened model represented by a line) are below the line.

classification to be very sensitive to the FUV and NUV errors, while $E(B-V)$ showed little change.

It quickly became apparent that many of the objects were incorrectly flagged as stars by the SDSS pipeline. The reddening expected around the NGP is not more than 0.07 magnitudes (Berdyugin & Teerikorpi 2002) yet we found many sources with a best-fitting $E(B-V)$ of 0.6 or even higher (Fig. 6). Much of this is due to confusion between stars and galaxies (Hutchings & Bianchi 2010) which may be removed using the object colours. For instance, main sequence stars will fall on the solid line in Fig. 7 ((NUV - u) versus (FUV - z)) with reddening pushing them above and to the right. Thus objects below the line cannot be stars.

As the original motivation in this work was to derive a 3-dimensional extinction map of our Galaxy, we have focused on the 14,360 of our objects which could actually be main sequence stars. However, we found that, even with this restricted data set, there were a significant number of sources with high reddening and thus not main sequence stars.

We extended our grid of models by including galaxy SEDs from the Kinney-Calzetti spectral atlas of galaxies (Calzetti, Kinney & Storchi-Bergmann 1994; Kinney et al. 1996). This spectral atlas includes galaxies of different morphological types from elliptical to late type spirals as well as for starburst galaxies. Starburst templates are provided for internal $E(B-V)$ ranging from 0.1 to 0.7 to which we have applied Weingartner & Draine (2001) extinction cross-sections for reddening within our Galaxy. Finally, we added a composite quasar spectrum with zero reddening from Vanden Berk et al. (2001), models for red giant stars of temperature ranging from

| Model | Temperature | log g | Reference |
|---|---|---|---|
| White dwarf | 32,300 K to 61,300 K | 7.5 to 8.0 | Bohlin, Colina & Finley (1995) |
| Main sequence stars | 3,500 K to 50,000 K | 4.0 to 5.0 | Castelli & Kurucz (2004) |
| Red giant | 3,500 K to 5,000 K | 1.0 to 3.0 | Castelli & Kurucz (2004) |
| Galaxy | - | - | Calzetti, Kinney & Storchi-Bergmann (1994) Kinney et al. (1996) |
| Quasar | - | - | Vanden Berk et al. (2001) |

Table 2. Summary of all the models used.

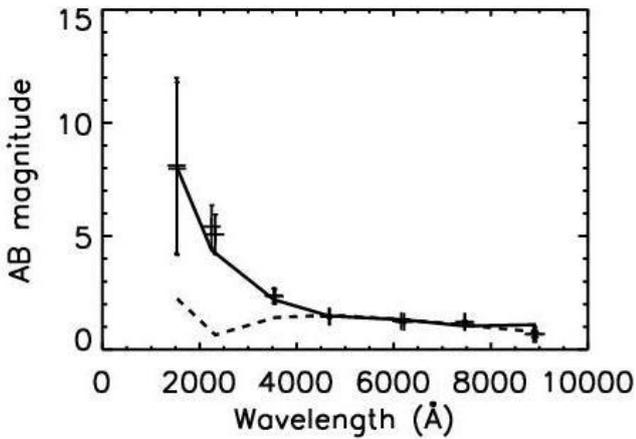

Figure 8. A sample plot showing the best model fit to a source at l = 118.83° & b = 76.46°, with a) Dashed line - a stellar SED from the Castelli and Kurucz atlas with *(B-V)* = 0.88, temperature = 50,000 K. (b) Solid line - a quasar SED from Vanden Berk et al. (2001) with *(B-V)* = 0.00.

3,500 K to 5,000 K with gravities 1.0 and 3.0 from Castelli & Kurucz (2004) and models for white dwarfs of temperature ranging from 32,300 K to 61,300 K with gravities spanning the range 7.5 to 8.0 from Bohlin, Colina & Finley (1995), to our model templates. A list of all the models used is summarised in Table 2. In addition to extinction, we also added a new free parameter, photometric redshift, to the galaxy and quasar models.

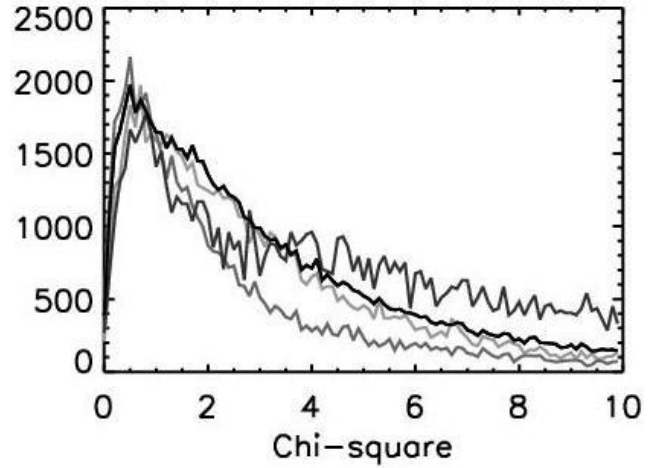

Figure 9. Chi-square distribution : for all objects (black), for quasars (dark grey), for starburst1 (grey) and for white dwarfs (light grey).

The photometric redshifts extend up to 2.0 and 6.0 respectively, in the case of galaxy and quasar models. We found that we were able to fit the observed SEDs much better with these extended models (Fig. 8) and were able to discriminate between stars and galaxies without a need for the anomalous Milky Way reddening.

The resulting overall reduced chi-square distribution is shown in Fig. 9, including that for object types that dominate the data set. We found the data set to be dominated mostly by quasars in the photometric redshift range from 0 to 2 (Fig. 10). The decline in the number of quasars between the redshift range of z ~ 2 to 3 was noted by Richards et al. (2002) as being due to a selection effect caused by the decrease in efficiency of the SDSS selection algorithm. The best fit distribution of the different classes is shown in Fig. 11. There are a number of different effects which may play role in this apparent distribution, the most important of which is the bias toward blue objects because of our insistence on *GALEX* fluxes.

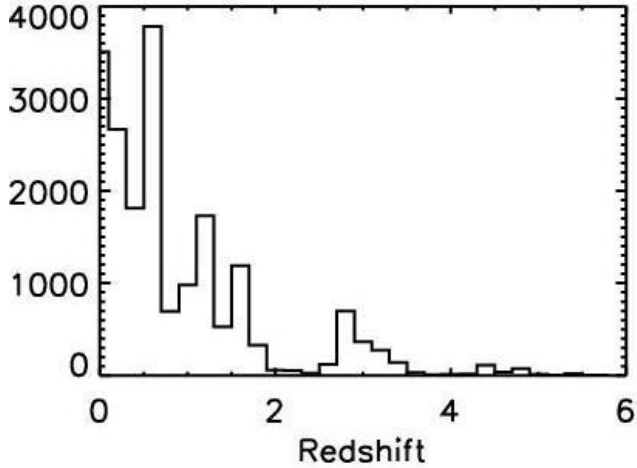

Figure 10. Histogram of redshift of quasars.

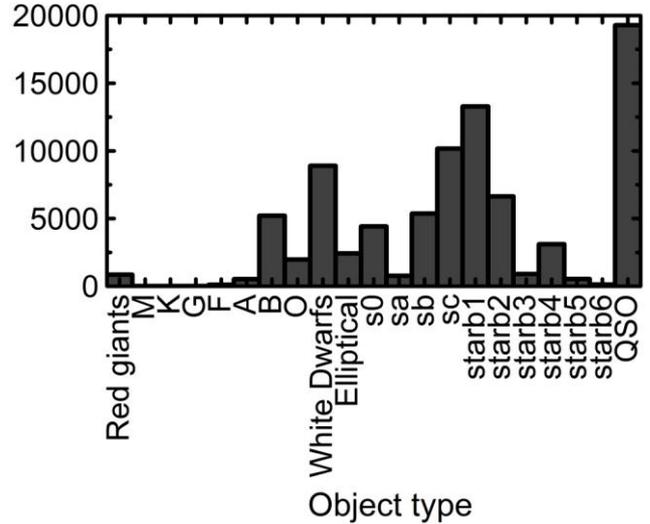

Figure 11. Histogram of objects in dataset by spectral type in case of stars and by morphological type in case of galaxies.

## 4 CONCLUSIONS

Although there has been considerable progress made in the analysis of survey data such as SDSS, the addition of UV data provides an invaluable lever arm in the diagnostics of Milky Way extinction and the identification of less common but interesting sources such as single hot white dwarfs, mixed-type binaries where a late-type star is paired with a hot white dwarf, etc.

We have found that purely colour based approaches cannot provide a clean separation between stars and galaxies and so have fit SED templates based on model stellar spectra (Calzetti & Kurucz 2004) and galaxy spectra (Kinney et al. 1996) to the observed SEDs. We have focused first on a limited sample of objects around the NGP where we expect little extinction. Our preliminary results show that matching model SEDs to the data provide effective discrimination between different classes of objects. As part of our fitting procedure, we classify the object, find its reddening and distance, and its photometric redshift. We found that many SDSS sources classified as stars are in fact galaxies and quasars.

Our template based fitting procedure allows us to easily add additional measurements, such as 2MASS data, and different object types and we now plan to apply this method to fit multi-wavelength data over the entire sky. As a next course of action, we will also be using the derived distances as an indicator to identify different objects. This is especially important with the large amount of data coming online over the next few years. One of our primary scientific goals will be to derive a 3-dimensional map of extinction in our Galaxy, necessary for interpreting any new data set.

**ACKNOWLEDGMENTS**

We would like to thank Ani Thakar and T. Sivarani for their help. This work is supported by the Department of Science and Technology (DST), Ministry of Science and Technology, Government of India, New Delhi, India under the grant No. SR/S2/HEP-011/2009.

Funding for SDSS-III has been provided by the Alfred P. Sloan Foundation, the Participating Institutions, the National Science Foundation, and the U.S. Department of Energy Office of Science. The SDSS-III web site is http://www.SDSS3.org/. SDSS-III is managed by the Astrophysical Research Consortium for the Participating Institutions of the SDSS-III Collaboration including the University of Arizona, the Brazilian Participation Group, Brookhaven National Laboratory, University of Cambridge, Carnegie Mellon University, University of Florida, the French Participation Group, the German Participation Group, Harvard University, the Instituto de Astrofisica de Canarias, the Michigan State/Notre Dame/JINA Participation Group, Johns Hopkins University, Lawrence Berkeley National Laboratory, Max Planck Institute for Astrophysics, Max Planck Institute for Extraterrestrial Physics, New Mexico State University, New York University, Ohio State University, Pennsylvania State University, University of Portsmouth, Princeton University, the Spanish Participation Group, University of Tokyo, University of Utah, Vanderbilt University, University of Virginia, University of Washington, and Yale University.

**REFERENCES**

Aihara H. et al., 2011, ApJS, 193, 29

Angelov T., 1996, Bull. Astron. Belgrade 154, 13

Bahcall J. N., 1986, ARA&A, 24, 575

Berdyugin A., Teerikorpi P., 2002, A&A, 384, 1050

Bertin E., Arnouts S., 1996, A&AS, 117, 393

Bohlin R. C., Colina L., Finley D.S., 1995, AJ, 110, 1316

Budavári T. et al., 2009, ApJ, 694, 1281

Calzetti D., Kinney A. L., Storchi-Bergmann T., 1994, ApJ, 429, 582

Castelli F., Kurucz R. L., 2004, arXiv:astro-ph/0405087v1

Eisenstein D. J. et al., 2011, AJ, 142, 72

Gunn J. E. et al., 1998, AJ, 116, 3040

Hutchings J. B., Bianchi L., 2010, AJ, 140, 1987

Kinney A. L., Calzetti D., Bohlin R. C., McQuade K., Storchi-Bergmann T., Schmitt H. R., 1996, ApJ, 467, 38

Lupton R. H., Ivezic Z., Gunn J. E., Knapp G., Strauss M. A., Yasuda N., 2002, Proc. SPIE, 4836, 350

Majewski S. R., 1993, ARA&A, 31, 575

Martin D. C. et al., 2005, ApJ, 619, L1

Morrissey P. et al., 2007, ApJS, 173, 682

Ravichandran S., Preethi K., Safonova M., Murthy J., 2013, Ap&SS, 344, 361

Richards G. T. et al., 2002, AJ, 123, 2945

Schlegel D. J., Finkbeiner D. P., Davis M., 1998, ApJ, 500, 525

Vanden Berk D. E. et al., 2001, AJ, 122, 549

Weingartner J. C., Draine B. T., 2001, ApJ, 548, 296

York D. G. et al., 2000, AJ, 120, 1579